\documentclass[letterpaper]{jpconf}

\usepackage{graphicx}
\usepackage{xspace}
\usepackage[square,sort&compress]{natbib}

\newcommand{\runi}{{\rm Run~I}\xspace}
\newcommand{\runii}{{\rm Run~II}\xspace}
\newcommand{\dzero}{D\O\xspace}

\newcommand{\mtop}{\ensuremath{m_{\rm top}}\xspace}
\newcommand{\ftop}{\ensuremath{f_{\rm top}}\xspace}
\newcommand{\psgn}{\ensuremath{P_{\rm sgn}}\xspace}
\newcommand{\pbkg}{\ensuremath{P_{\rm bkg}}\xspace}
\newcommand{\pevt}{\ensuremath{P_{\rm evt}}\xspace}
\newcommand{\etmiss}{\ensuremath{E \kern-0.6em\slash_{T}}\xspace}
\newcommand{\etmissxobs}{\ensuremath{E \kern-0.6em\slash_{x}^{obs}}\xspace}
\newcommand{\etmissyobs}{\ensuremath{E \kern-0.6em\slash_{y}^{obs}}\xspace}
\newcommand{\etmissx}{\ensuremath{E \kern-0.6em\slash_{x}}\xspace}
\newcommand{\etmissy}{\ensuremath{E \kern-0.6em\slash_{y}}\xspace}
\newcommand{\etmissxsmall}{\ensuremath{E \kern-0.4em\slash_{x}}\xspace}
\newcommand{\etmissysmall}{\ensuremath{E \kern-0.4em\slash_{y}}\xspace}
\newcommand{\ifb}{\ensuremath{\rm fb^{-1}}\xspace}
\newcommand{\keV}{\ensuremath{\mathrm{ke\kern-0.1em V}}\xspace}
\newcommand{\MeV}{\ensuremath{\mathrm{Me\kern-0.1em V}}\xspace}
\newcommand{\GeV}{\ensuremath{\mathrm{Ge\kern-0.1em V}}\xspace}
\newcommand{\GeVc}{\ensuremath{\mathrm{Ge\kern-0.1em V}/c}\xspace}
\newcommand{\GeVcc}{\ensuremath{\mathrm{Ge\kern-0.1em
      V}/c^{2}}\xspace}
\newcommand{\TeV}{\ensuremath{\mathrm{Te\kern-0.1em V}}\xspace}

\newcommand{\W}{\ensuremath{W}\xspace}
\newcommand{\Zjets}{\ensuremath{Z+{\rm jets}}\xspace}
\newcommand{\WWjets}{\ensuremath{WW+{\rm jets}}\xspace}
\newcommand{\WZjets}{\ensuremath{WZ+{\rm jets}}\xspace}
\newcommand{\ZZjets}{\ensuremath{ZZ+{\rm jets}}\xspace}
\newcommand{\Wjets}{\ensuremath{W+{\rm jets}}\xspace}
\newcommand{\msbar}{\ensuremath{\overline{\rm MS}}\xspace}

\newcommand{\Eref}[1]{(\ref{#1})}

\begin{document}

\title{Top quark mass measurements at the \dzero experiment}

\author{Alexander Grohsjean} 

\address{Ludwig-Maximilians-Universit\"at M\"unchen}

\begin{abstract}
The most recent measurements of the top quark mass at the \dzero
experiment are summarized. Different techniques and final states are
used and the top quark mass is determined to be $\mtop=172.8\pm1.6({\rm
  stat+syst})\GeVcc$. In addition, a new, indirect measurement
comparing the measured cross section to theoretical calculations 
is discussed. Both, the direct and the indirect measurement of the top
quark mass are in good agreement. 
\end{abstract}

\section{Introduction}
The top quark is the heaviest of all known fundamental particles.
Due to its
high mass, the Yukawa coupling of the top quark to the Higgs boson 
is close to unity suggesting that it may play a special role in
electroweak symmetry breaking~\cite{bib:HASHIEXTDIM}. Moreover, precise
measurements of its mass constrain the mass 
of the yet-unobserved Higgs boson through radiative corrections to the
weak coupling and allow for a restriction of possible extension to
the Standard Model~\cite{bib:HEINETOPIMPACT}. 

At the Tevatron collider with a centre-of-mass energy of 1.96~\TeV, 
85\% of the top quark pairs are produced in quark-antiquark annihilation;
15\% originate from gluon fusion. Both, top and antitop are
predicted to decay almost exclusively to a \W boson and a bottom
quark. According to the number of hadronic \W decays, top quark events
are classified into all-jets, lepton+jets and dilepton events. 

The lepton+jets channel is characterized by four jets, one
isolated, energetic charged lepton\footnote{Here and
  throughout this article, lepton refers to electron or muon.}
 and missing transverse energy. With 30\%, the branching fraction of
 the lepton+jets channel is about seven times larger than the one of
 the dilepton channel whereas 
the signal to background ratio is about three times smaller. The main 
background in this final state comes from \Wjets events.      
Instrumental background arises from events in which a jet is
misidentified as an electron and events with heavy hadrons 
that decay into leptons which pass the isolation requirements.

The topology of the dilepton channel is described by two
jets, two isolated, energetic charged leptons and
significant missing transverse energy from the undetected neutrinos.    
The dilepton channel has the smallest branching ratio but also  
the smallest background contamination. The main backgrounds
are \Zjets and diboson events (\WWjets, \WZjets, \ZZjets) as well as
instrumental background as characterized above.

At the \dzero experiment, two different techniques are
used to measure the top quark mass in the dilepton final state: the
Neutrino Weighting method~\cite{bib:LLNUWEIGHT08} and the Matrix
Element method~\cite{bib:LLME08}. The latter one is also used in the
lepton+jets final state where the mass of the hadronically decaying
\W~boson yields an additional constrain to determine the jet
energy scale\footnote{In this context, jet energy scale refers to an
  additional scalar correction factor on top of the nominal jet energy
correction applied at the \dzero experiment.} 
and the top quark mass simultaneously and to achieve a
significant reduction of the main systematic uncertainty~\cite{bib:LJME08}.      
Complementary, the top quark mass can be extracted from the
measured top pair production cross section comparing it to theoretical
calculations~\cite{bib:MASSFROMXSEC}. 

\section{The Neutrino Weighting Method}

The Neutrino Weighting method is a template based method and has
already been used during the \runi period of the Tevatron accelerator.
For each event, the neutrino momenta are calculated assuming
a certain top mass and different neutrino pseudorapidities. 
A weight $w$ is assigned according to the agreement of the calculated
sum of the neutrino momenta $\sum_{i=1}^2
p^{\nu_{i}}_x$, $\sum_{i=1}^2
p^{\nu_{i}}_y$, and the measured missing
transverse momentum components \etmissx,\etmissy in the event, given by
\begin{equation}
\label{eq:nuweight}
       w = \exp (\frac{-(\etmissx
	 - \sum_{i=1}^2 p^{\nu_{i}}_x)^2}{2\sigma_{\etmissxsmall}^2 }) 
       \exp (\frac{-(\etmissy
	 - \sum_{i=1}^2 p^{\nu_{i}}_y)^2}{2\sigma_{\etmissysmall}^2 })
       \,,
\end{equation}
where $\sigma_{\etmissxsmall},\sigma_{\etmissysmall}$ denote the
resolution of the missing energy measurement.
This process is repeated many times varying the jet and lepton
energies within their experimental resolution. Next, 
signal probability distributions as a function of the top quark mass,
the mean and the RMS of the weight distributions are derived using Monte Carlo
simulated signal events of different top quark masses. To reduce the
bias from background, background probability
distributions as a function of
the mean and the RMS of the weight distributions are determined
accordingly using simulated \Zjets and diboson events.  
Finally, a likelihood function that consists of three terms is built to
measure the top quark mass. The first term
accounts for the agreement of the expected number of signal and
background Monte Carlo events to the one in the data sample, the
second for the agreement of the background events with the prediction,
and the third for the agreement of the data with the signal and
background probability distributions. The method is calibrated using
ensemble tests. 

For the measurement, events in the dilepton final state are selected
requiring two isolated leptons, or alternatively one isolated lepton and one
isolated, charged particle track. 
The data set analyzed corresponds
to an integrated luminosity of about 1~\ifb collected between April
2002 and February 2006. A kinematic selection
that reduces the contamination from background to about 20\% is applied.
The top quark mass is extracted from 82 top pair candidate events
yielding   
\begin{equation}
\mtop=176.0\pm5.3({\rm stat})\pm2.0({\rm syst})~\GeVcc .
\end{equation}
The main systematic uncertainties on this measurement come from
the energy scale of the jets, the modeling of the simulated signal
events and the fragmentation of the jets from $b$ quarks. 

\section{The Matrix Element Method}
The Matrix Element method was developed at the \dzero experiment 
during the \runi period of the Tevatron to extract the top
quark mass with high precision from a sample of low
statistics~\cite{bib:MERunINature}.  
The Matrix Element method yields the single most precise
measurement of the top quark mass.  

The probability for a final state $y = (p_1,\dots,p_6)$ of $6$
partons of four-momenta $p_i, i=1,\dots,6$  to originate from the
signal process under the assumption of a certain top mass \mtop
is given by 
\begin{equation}
\label{eq:psgn}
\psgn(x;\mtop) = \frac{1}{\sigma_{\rm obs}(\mtop)} \int
~{\rm d}\epsilon_1~{\rm d}\epsilon_2~f_{\rm PDF}(\epsilon_1)~f_{\rm PDF}(\epsilon_2)
~~\frac{(2 \pi)^{4} \left|M(y)\right|^{2}}
       {\epsilon_1\epsilon_2 s}
 ~     {\rm d}\Phi_{6}
 ~~ W(x,y) \ , 
\end{equation}
where $\epsilon_1$, $\epsilon_2$ denote the energy fraction of the
incoming quarks from the protons and antiprotons, $f_{\rm PDF}$ the
parton distribution function, $s$, the centre-of-mass
energy squared, $M(y)$, the
leading-order matrix element~\cite{bib:MAHLON} and ${\rm d}\Phi_{6}$,
an element of the 6-body phase space.   
The finite detector resolution is taken into account using a transfer
function $W(x,y)$ that describes the probability of a partonic final
state $y$ to be measured as $x = (\tilde{p}_1,\dots,\tilde{p}_n)$ in the
detector. 
The signal probability is normalized with the observable cross section
$\sigma_{\rm obs}$.

In a similar way, for each event the probability to arise from
background is calculated. Taking the huge amount of computing time
into account, only the leading source of background is considered,
i.e. \Zjets probabilities in the dilepton case, \Wjets probabilities
in the lepton+jets channel. In a last step, both probabilities, signal
and background, are combined to an event probability taking the signal 
fraction \ftop into account   
 \begin{equation}
  \pevt(x;\mtop) = \ftop\cdot\psgn(x;\mtop)+(1-\ftop)\cdot\pbkg(x) \, ,
\end{equation}
and the top quark mass is extracted from a likelihood fit of the
product of the event-by-event probabilities.  
To calibrate the method and correct for any bias, Monte
Carlo simulated events are used to perform ensemble tests. 
\subsection{Dilepton Channel}
While only events with exactly one electron and exactly one muon 
are taken into account so far, the measurement makes use of the full
\runii data set recorded between April 2002 and May 2008. This
corresponds to an integrated luminosity of 2.8~\ifb. To reduce the
fraction of background events, a kinematic selection is
applied leaving 107 top pair candidate events with a signal fraction
of about 80\%. The top quark mass is
measured  to be  
\begin{equation}
\mtop=172.9\pm3.6({\rm stat})\pm2.3({\rm syst})~\GeVcc .
\end{equation}
The dominant sources of systematic uncertainties are  
jet uncertainties, such as their energy scale and resolution.
With an statistical uncertainty of $3.6$~GeV, this measurement has the
smallest uncertainty of all top mass measurements performed in the
dilepton channel at the \dzero experiment. 

\subsection{Lepton+Jets Channel}
As stated above, the largest systematic uncertainty on the top quark
mass arises from the jet energy scale. Since one of the \W boson decays
hadronically in the lepton+jets channel, the well known \W mass 
can be used to constrain the jet energies. An additional scale factor 
is introduced in Eq.~\Eref{eq:psgn} and both, the top
quark mass and the jet energy scale are measured simultaneously.

In the lepton+jets channel, an integrated luminosity of 2.2~\ifb is
used. A kinematic selection leaves 491 data events in total, the
signal fraction is about 40\%. The top quark mass in this sample
is measured to be 
\begin{equation}
\mtop=172.2\pm1.0({\rm stat+JES})\pm1.4({\rm syst})~\GeVcc .
\end{equation}   
The systematic uncertainty is dominated by the uncertainty on the
difference between the nominal inclusive response and the
response of jets from $b$ quarks in the calorimeter. 

\section{Top Quark Mass From Production Cross Section}

Two different schemes are commonly used to define the top quark mass: 
the \msbar and the pole mass definition. Direct top quark mass
measurements are based on leading-order Monte Carlo generators 
where the top quark is described by a Breit-Wigner resonance and    
the measured mass corresponds only approximately to a pole mass. 
Since the top quark pair production cross section depends on the top quark
mass, a measurement with a well-defined and renormalization-scheme
independent mass definition can be realized comparing  
the results from cross section measurements 
to fully inclusive calculations in higher-order QCD. Here,
the pole mass definition is applied and the extracted value can be
unambiguously interpreted. 
Comparing the combined measured cross sections from the lepton+jets,
dilepton and lepton+tau channel to the NLO+NNLL calculations from
S.~Moch~{\it et~al.}~\cite{bib:moch-2008-78} yields a top quark mass of
$169.6^{+5.4}_{-5.5}~\GeVcc$, while 
the comparison with the NLO+NLL calculations from M.~Cacciari~{\it
  et~al.}~\cite{bib:cacciari-2008-0809} yields
$167.8^{+5.7}_{-5.7}~\GeVcc$. Both results are in good 
agreement with the direct measurements. 


\section{Conclusion}
The top quark mass has been measured at the \dzero experiment 
in different final states using different techniques. 
In the dilepton channel, the largest systematic uncertainty arises
from the jet energy scale. This has been addressed in the lepton+jets
channel with a 
simultaneous measurement of the top quark mass and the jet energy
scale which is well constrained by the mass of the hadronically
decaying \W boson. Combining all measurements with the Best Linear
Unbiased Estimate (BLUE)~\cite{bib:COMBI08}, the top quark mass is
determined at the \dzero experiment to be 
\begin{equation}
\mtop=172.8\pm1.6({\rm stat+syst})~\GeVcc.
\end{equation}
The top quark mass has also been extracted from cross section
measurements comparing them to fully inclusive calculations in
higher-order QCD. Both measurements, direct and indirect, 
are in good agreement with each other. 

\ack
We thank the staffs at Fermilab and collaborating institutions, 
and acknowledge support from the 
DOE and NSF (USA);
CEA and CNRS/IN2P3 (France);
FASI, Rosatom and RFBR (Russia);
CNPq, FAPERJ, FAPESP and FUNDUNESP (Brazil);
DAE and DST (India);
Colciencias (Colombia);
CONACyT (Mexico);
KRF and KOSEF (Korea);
CONICET and UBACyT (Argentina);
FOM (The Netherlands);
STFC and the Royal Society (United Kingdom);
MSMT and GACR (Czech Republic);
CRC Program, CFI, NSERC and WestGrid Project (Canada);
BMBF and DFG (Germany);
SFI (Ireland);
The Swedish Research Council (Sweden);
CAS and CNSF (China);
and the Alexander von Humboldt Foundation (Germany).


\bibliography{llme}

\end{document}